# Polytype control of spin qubits in silicon carbide

Abram L. Falk[1], Bob B. Buckley[1], Greg Calusine[1], William F. Koehl[1], Viatcheslav V. Dobrovitski[2], Alberto Politi[1], Christian A. Zorman[3], Philip X.-L. Feng[3], David D. Awschalom[1]

1. Center for Spintronics and Quantum Computation, University of California, Santa Barbara, California 93106, USA

2. Ames Laboratory and Iowa State University, Ames, Iowa 50011, USA

3. Department of Electrical Engineering, Case School of Engineering, Case Western Reserve University, Cleveland, OH 44016, USA

## Abstract

Crystal defects can confine isolated electronic spins and are promising candidates for solid-state quantum information. Alongside research focusing on nitrogen vacancy centers in diamond, an alternative strategy seeks to identify new spin systems with an expanded set of technological capabilities, a materials driven approach that could ultimately lead to "designer" spins with tailored properties. Here, we show that the 4H, 6H and 3C polytypes of SiC all host coherent and optically addressable defect spin states, including spins in all three with room-temperature quantum coherence. The prevalence of this spin coherence shows that crystal polymorphism can be a degree of freedom for engineering spin qubits. Long spin coherence times allow us to use double electron-electron resonance to measure magnetic dipole interactions between spin ensembles in inequivalent lattice sites of the same crystal. Together with the distinct optical and spin transition energies of such inequivalent spins, these interactions provide a route to dipole-coupled networks of separately addressable spins.

## Introduction

The search for coherently addressable spin states[1] in technologically important materials is a promising direction for solid-state quantum information science. Silicon carbide, a particularly suitable target[2-4], is not a single material but a collection of about 250 known polytypes. Each polytype is a binary tetrahedral crystal built from the same two-dimensional layers of silicon and carbon atoms, but different stacking sequences give each its own crystal structure, set of physical properties, and array of applications. 4H- and 6H-SiC, the most common hexagonal polytypes, are used for power and opto-electronics, and as growth substrates for graphene[5] and gallium nitride[6]. The cubic 3C-SiC can be grown epitaxially on silicon[7] and is often used in micromechanics[8]. Driven by computational[2,9-11], electron paramagnetic resonance[9,10,12-17] and optical studies[3,12,13,18-22], research into defect-based spins in SiC has also led to their increasing appreciation as candidate systems for quantum control.

Our results demonstrate that despite their varying optical, electronic, and structural properties, the three most common SiC polytypes all exhibit optically addressable spin states with long coherence times. These spins are localized electronic states bound to neutral divacancies[9] and related defects. Increasingly complex polytypes of SiC can host an increasing number of inequivalent defect sites–for instance, there are $n$ inequivalent divacancy sites in $n$H-SiC.

We measure magnetic dipole-dipole interactions between ensembles of inequivalent spins and use these interactions to infer an estimate for the degree of optical spin polarization, 35-60%, depending on the defect species. These high polarization values are an important parameter for optically addressed spin control. Moreover, since these inequivalent defect sites are separately addressable through distinct optical and spin transition energies, extending our results to the single-spin limit could lead to quantum networks of dipole-coupled spins.

## Results

### 1. Optically detected spin states in SiC

To generate defect ensembles in SiC, we began with semi-insulating (SI), n-type, and undoped SiC substrates. In substrates with a low intrinsic defect concentration, we then used carbon ion implantation followed by an annealing process[12] designed to join vacancies into complexes (see Methods and Supplementary Note 1). The 3C-SiC samples consist of single and poly-crystalline epitaxial films grown on silicon substrates, while the 4H- and 6H-SiC substrates are bulk single crystals.

Our optically detected magnetic resonance (ODMR) measurements show that all three measured polytypes host a number of optically addressable defect spins. These ODMR measurements rely on spin-dependent optical cycles both to polarize spins with laser illumination and to measure those spin states through changes in the photoluminescence (PL) intensity. We focus on defect optical transitions with zero-phonon lines in the 1.08 – 1.2 eV range[12,23]. These can be observed as peaks in the PL spectrum when the samples are subject to 1.27 – 1.45 eV laser excitation (Figure 1a). In addition to these sharp peaks, much of the PL from these defects is emitted in broad phonon sidebands at lower energies, which we also collect.

The ODMR spectra (Figs. 1b-d) are obtained by measuring the fractional change in PL intensity ($\Delta I_{PL}/I_{PL}$) under continuous wave laser illumination as a function of both an applied out-of-plane DC magnetic field ($B$) and the frequency ($f$) of an applied radiofrequency (RF) magnetic field. Spin flips produce a $\Delta I_{PL}/I_{PL}$ signature and occur when $f$ is resonant with one of the defects' spin transitions, which can all be tuned by varying $B$. The large number of observed ODMR lines demonstrates the versatility of SiC as a host for optically addressable spin states.

In each polytype, wavelength-resolved ODMR measurements associate the various ODMR features with specific PL lines (see refs 3,13 for 4H-SiC and Supplementary Figure S1 for 6H- and 3C-SiC). Some of the defects in 4H-SiC (PL1-PL4) have been identified as spin-1 neutral divacancies[9]. The defects responsible for the other spin transitions observed (compiled in Supplementary Table S1) have similar spin and optical properties to the neutral divacancies but have not been conclusively identified.

### 2. Spin coherence at cryogenic and room temperatures

Long-lived spin coherence, an important prerequisite for quantum information and sensing technologies, is a general feature of spins in all three polytypes. Our coherence measurements are based on standard pulsed magnetic resonance techniques including Rabi, Ramsey, Hahn echo, Carl-Purcell-Meiboom-Gill (CPMG) (Figure 2) and spin relaxation (Supplementary Figure S2) sequences. At 20 K, the spin relaxation times range from 8 to 24 ms. The Hahn-echo coherence times ($T_2$) range from 10 μs to 360 μs at 20K, depending on the substrate, with significant dependence on implantation dose and substrate doping type. The longest $T_2$ times we measured were in native neutral divacancies in 4H-SiC that were generated during crystal growth.

All three polytypes exhibit defects whose spin coherence persists up to room temperature (Figure 3 and Supplementary Figures S3-S5). In as-grown 4H-SiC, one neutral divacancy line (PL3) persists up to room temperature as well as three other ODMR lines (PL5-PL7) of unknown origin, all with $T_2$=50 +/- 10 µs. Polycrystalline 3C-SiC also exhibits a state with room-temperature spin coherence, although similar states with the same zero phonon line and ODMR transition in certain other 3C-SiC substrates that we measured did not. In 6H-SiC, the SI and n-type substrates have the same ODMR lines at 20K, but their room-temperature ODMR signatures are substantially different from each other (Figure 3b), with several additional ODMR lines in the n-type substrate that do not appear at 20K or in the SI substrate. The presence of these coherent spin states at room temperature is a particularly promising result for spin-based sensing[24] with SiC.

## 3. Coherent spin interactions

Ultimately, many spin-based quantum technologies will require not only separately addressable spins with long coherence times, but also a means of coupling these spins together. Patterned ion implantation has generated individual and strongly coupled nitrogen-vacancy (NV) centers in diamond[25]. We have also patterned spin ensembles in SiC, using ion implantation through poly-(methyl-methacrylate) (PMMA) apertures (Figure 4a and Supplementary Figure S6). Though these are not individual spins, this patterning demonstration shows promise for spatially engineering SiC defects. Since optical wavelengths significantly exceed the length scale required for strong magnetic coupling between single dipoles (< 30 nm for diamond NV centers[25]), scaling up a dipole-coupled spin network is a significant challenge. Silicon carbide defects in inequivalent lattice sites have distinct RF and optical transition energies, giving complex polytypes of SiC with many inequivalent defect species the possibility of hosting many separately addressable spins (Fig. 4b) in a single confocal volume (Fig. 4b).

As a step towards independently addressable dipole-coupled spins, we measure dipole-dipole spin interactions between inequivalent defect ensembles. The study of these interactions also provides valuable information about the spin density and optical polarization in these defect states. Our measurements use double electron-electron resonance**[26]** (DEER) to flip the spin of one spin ensemble (the "drive" species) while the resulting change in the Larmor precession rate of another ensemble (the "sense" species) is measured. We focused on 6H-SiC for these measurements, which when implanted, had higher spin densities than in our 4H-SiC substrates, and higher DEER coupling strengths.

The change in precession rate ($\Delta f$), a measure of the average dipole coupling strength, is experimentally observed as an additional phase ($\Delta \theta_\text{free}$) acquired by the sense species over the free precession segment of a Hahn echo measurement (Figure 5a). These parameters are related by $\Delta \theta_\text{free} = 2\pi \Delta f (2 t_\text{pulse})$, where $t_\text{pulse}$ is the delay of the drive pulses relative to the center of the Hahn echo sequence. This pulse sequence is designed to refocus the sense spins due to all magnetic fields except those due to drive species spin flips. When we drive Rabi oscillations on the drive species, we simultaneously observe DEER oscillations in $\Delta \theta_\text{free}$ of the sense spins (Figure 5b).

In order to measure both $\Delta f$ and the decoherence rate of the sense spins due to drive spin flips ($\tau$), we vary both $t_\text{pulse}$ and the phase of the final $\pi/2$ pulse in the sense spin Hahn echo sequence ($\theta_\text{Hahn}$). The resulting data (Fig. 5c: left) are well fit by:

$$\text{DEER signal} = \cos(2\pi \Delta f (2 t_\text{pulse}) + \theta_\text{Hahn}) e^{-2 t_\text{pulse}/\tau}. \tag{1}$$

For the data at $\theta_\text{Hahn} = \pi/2$, the coherent coupling term becomes $\sin(2\pi \Delta f (2 t_\text{pulse}))$ in equation (1), providing a sensitive measure of $\Delta f$, while the data at $\theta_\text{Hahn}=0$ and $\theta_\text{Hahn}=\pi$ are dominated by the decoherence term, giving a more accurate measure of $\tau$. The globally fitted values of $\Delta f$ for various drive spin ensemble species (Figure 5d and Supplementary Figure S7) show that the c-axis oriented spins (QL1, QL2, and QL6) exhibit $\Delta f$ values in the single kHz range. The lower symmetry of the basal-oriented spins (QL3-QL5) results in eigenstates with smaller magnetic moments, reducing $\Delta f$ for these species (Supplementary Note 2).

We also repeated this experiment with the pulse sequence shown in Figure 5a modified by the addition of a depolarizing π/2 pulse, which is applied to the drive spins before the sense spins' Hahn echo sequence. In this case, because the π pulses applied to the drive spin population no longer cause a net change in magnetization, Δ$f$ vanishes (Figure 5c: right). Additionally, when the drive spins are depolarized, τ increases slightly. When QL1 is the sense species and QL2 is the drive species, τ changes from 89 ± 3 μs (polarized QL2) to 96 ± 3 μs (unpolarized QL2).

## Discussion

The decoherence characterized by τ, known as instantaneous diffusion, arises from two effects. The first (and dominant) of these is microscopically inhomogeneous dipole coupling strengths between randomly located individual spins, resulting in a distribution of coupling strengths, whose average is Δ$f$. This distribution causes ensemble dephasing of the sense species when the drive species is flipped. The second effect is a macroscopically inhomogeneous magnetization field due to the spatial structure of the optically polarized volume of spins. The result of this effect is a slightly longer τ when the spin bath is depolarized, consistent with the data. The analysis of these data is complicated by the dynamics of discretely interacting spins in a polarized spin bath (Supplementary Note 3). Nevertheless, the standard quasi-static statistical model relating instantaneous diffusion to spin density[26,27] provides a guide for analyzing the DEER results.

Based on this model, we use τ for the unpolarized spin bath to infer that the spin densities of the three c-axis spin species (QL1, QL2, and QL6) range from 7-11 ×10$^{15}$ spins/cm$^3$. Because Δ$f$ is proportional to the magnetic field generated by the drive spin ensemble, which in turn is proportional to the product of spin density and optical polarization, we can use the measured Δ$f$ and the calculated spin density to infer the degree of optical spin polarization. For the three c-axis spins, our DEER results lead to a high average optical spin polarization that ranges from 35 to 60%, depending on the defect species. Due to the inhomogeneous spatial profile of the optical illumination and collection areas in our measurements, the full optical polarization is likely to be even higher.

SiC spins are compelling analogues of diamond NV centers, with complementary properties and many unique prospects. Our demonstration that crystal polymorphism can be used to engineer new spin centers relies on SiC being a polymorphic material, a degree of freedom that is unavailable in diamond. In the future, established doping and epitaxial growth processes in SiC could lead to electronic interfaces with defect spins embedded in transistors and optoelectronic devices. Furthermore, due to its availability as a single-crystalline epitaxial film on silicon, the 3C polytype provides an excellent platform for hybrid quantum systems with photonic[28] and mechanical[29] degrees of freedom. Combining this sophisticated semiconductor technology with the versatility of coherent spin control in SiC stands to be an exciting route for solid-state quantum information.

## Methods

1. **Generation of defects**

The 4H-SiC and 6H-SiC substrates used in this work include: a. n-type 4H-SiC b. high-purity SI 4H-SiC, c. n-type 6H-SiC, and d. SI 6H-SiC. Substrates a, b, and c were purchased from Cree, Inc, while d was purchased from II-VI, Inc. The SI 4H-SiC contains a significant density of neutral divacancy spins as grown, and the data in Figures 1, 2a and 3 used that material without modification. Defects in the other substrates were generated by an ion implantation process consisting of 190 keV $^{12}$C ion implantations at doses of 10$^{11}$, 10$^{12}$, and 10$^{13}$ cm$^{-2}$, with a 7 degree tilt to minimize ion channeling effects.

After ion implantation, the samples were annealed at 900 °C for 30 minutes in Ar, a process designed to allow vacancies to diffuse and aggregate into pairs and vacancy complexes[12]. We estimate a 5% creation efficiency of fluorescent defects, defined as the number of created defects per implanted $^{12}$C ion at 190 keV (Supplementary Note 1). For the n-type substrates, ion implantation with $^{12}$C can compensate the n-type doping. Throughout the text, our labeling of substrate doping types refers to the as-grown substrates, not to the semiconductor characteristic after implantation.

The 3C-SiC substrates used in this work consisted of [100] oriented epitaxial layers grown on [100] silicon substrates. The 3C-SiC substrates measured in Figs. 1 and 2 were 3.85 μm thick films and were obtained from Novasic. The substrates measured in Fig. 3 were grown at Case Western Reserve University and consisted of 1.5-2 μm thick polycrystalline films. Neither film was intentionally doped but both showed n-type behavior. All samples were implanted with 190 keV $^{12}$C at doses of $10^{12}$ or $10^{13}$ cm$^{-2}$ with a 7 degrees tilt. The samples were the annealed at 750 °C for 30 minutes in Ar.

The patterned defects in Fig. 4a were generated by ion implantation at a lower energy (10 keV $^{12}$C ions at a dose of $10^{13}$ cm$^{-2}$), which makes the 170 nm-thick PMMA a more effective ion mask. The PMMA mask used for the sample in Fig. 4b consisted of 50 nm apertures patterned by electron beam lithography.

## 2. Optically detected magnetic resonance

For our ODMR measurements, the laser excitation was higher energy than the defects' zero-phonon lines, within their absorption sidebands. For the 4H- and 6H- substrates in Figs. 1-4, the laser energy was 1.45 eV (853 nm), with 16 mW of power reaching the sample. For the 3C-SiC data in Figs. 1d and 2b, the laser energy was 1.33 eV (930 nm), and the power was 23 mW at the sample. For the double resonance data in Fig. 5, the laser energy was 1.27 eV (975 nm), with 60 mW reaching the sample. The laser excitation was gated with acousto-optical modulators and the fluorescence was collected using one of four detectors: a) a Thorlabs Femtowatt InGaAs photoreceiver (PDF10C), b) a Newport InGaAs photoreceiver (2011-FS), c) a Princeton instruments liquid-nitrogen cooled InGaAs camera attached to an Acton Spectrometer (2300i), and d) a Scontel superconducting detector (LTD 24/30-008).

The samples were mounted on top of 0.5 – 2 mm RF strip lines[3]. For the 3C-SiC, ring-shaped RF waveguides fabricated on chip were also used[3]. These sample/waveguide assemblies were then mounted in optical cryostats with RF access. The RF signals were generated by two signal generators (Agilent E8257C or Rohde & Schwarz SM300 vector source) whose outputs were gated using RF switches (MiniCircuits ZASWA-2-50DR+) for pulsed experiments. These signals were then combined, amplified to peak powers as high as 25W (Amplifier Research 25S1G4A and Mini-Circuits ZHL-30W-252-S+), and sent to wiring in the cryostat connected to the waveguides and striplines. The RF and optical pulses were gated with pulse patterns generated by either a digital delay generator (Stanford Research Systems DG645), Pulse Pattern Generator (Agilent 81110A) or arbitrary waveform generator (Tektronix AWG520). The phase of the Rohde & Schwartz signal was also controlled by the AWG520.

The ODMR measurements in this paper were all taken using lock-in techniques, in which an RF pulse was alternatively gated on and off (Figs. 1 and 3a-c) or the phase of one of the pulses was alternatively gated by 180° using IQ modulation (Figs. 2, 3d-e, and 5). We used a 20 Hz software lock-in technique[3] (Figs. 1-3) and a hardware lock-in at frequencies up to 200 kHz (Fig. 5) to accommodate the bandwidths of the Thorlabs and Newport photoreceivers respectively.

Because the PL spectra of the various divacancy orientations have overlapping phonon sidebands, these measurements collected $I_{PL}$ from all defect orientations at once. This procedure reduced the normalized $\Delta I_{PL}/ I_{PL}$ signal but prevented defect PL from being rejected. The measured $\Delta I_{PL}/ I_{PL}$ values are additionally reduced from their ideal values both by extra fluorescence from the SiC samples (notably Vanadium impurities in the SI 6H-SiC) and by long fluorescence collection times. In order to achieve high optical spin polarization, we used long optical pulses (20 – 100 μs) in our measurement cycles. Because the timescale of optical polarization was faster than the bandwidth of most of our detectors, we did not gate the fluorescence collection. Therefore, much of the PL collected was from defect spins already polarized earlier in the pulse. Fast and gated fluorescence detection would lead to significantly higher $\Delta I_{PL}$ values.

**Supplementary Information** is available below and in the peer-reviewed version of this paper: A. L. Falk *et al.*, *Nat. Commun.,* **4**, 1819 (2013), doi: 10.1038/ncomms2854.

**Acknowledgments** We are grateful to Ádám Gali, David Christle, and David Toyli for reviewing our manuscript and for helpful discussions. This work was supported by the Air Force Office of Scientific Research (AFOSR).

**Author Contributions** All authors developed the concept and designed the experiments. The data for the hexagonal polytypes were taken by A. L. F., B. B. B., and W. F. K., and the 3C-SiC studies were performed by G. C. and A. P. The epitaxial 3C-SiC were grown by C. A. Z. and P. X.-L. F. All authors contributed to the analysis and manuscript.

**Corresponding author** Correspondence should be addressed to D. D. Awschalom at awsch@physics.ucsb.edu.

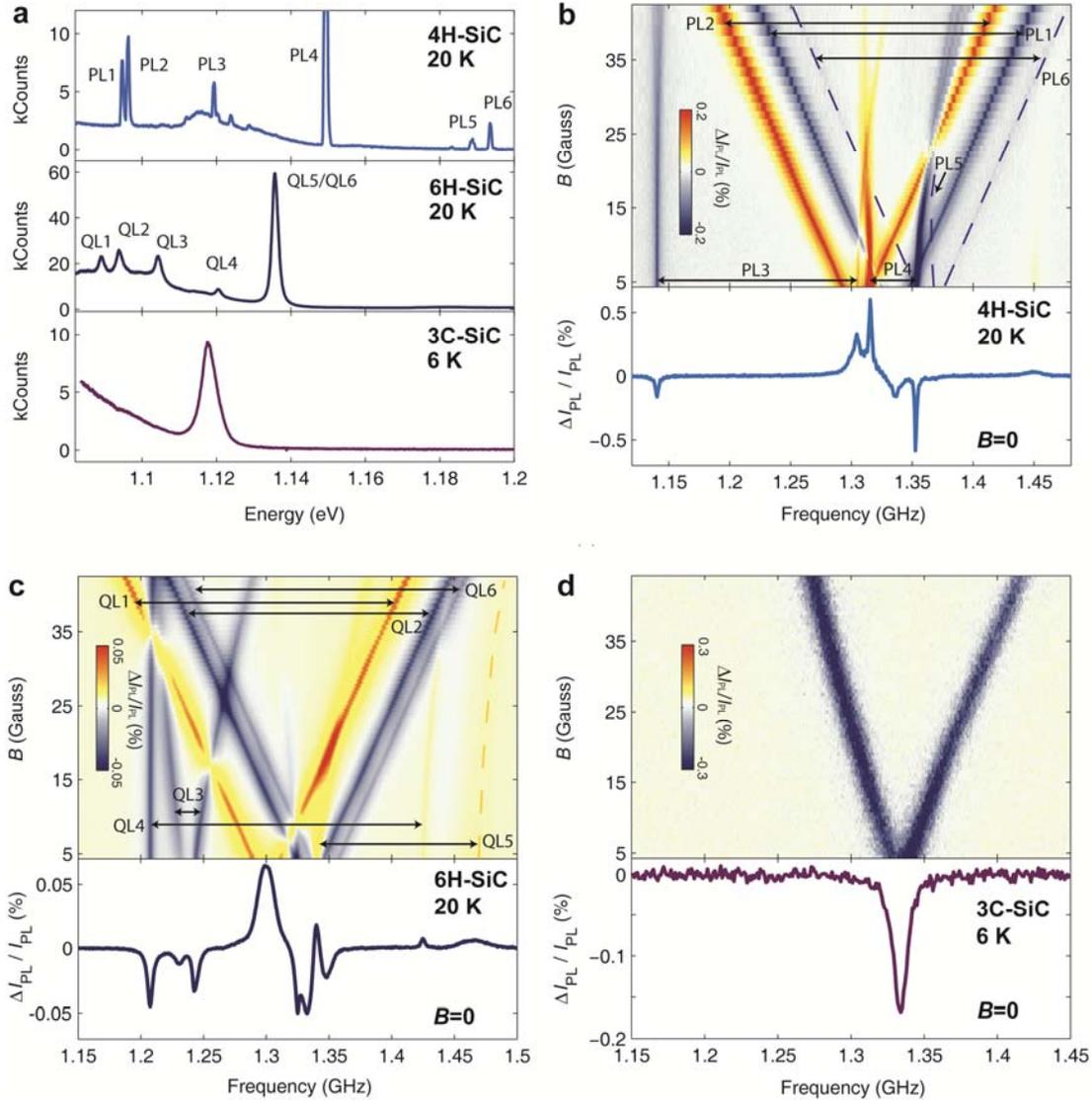

**Figure 1. Optical and spin transition spectra in the three most common SiC polytypes. a.** Optical spectrum of as-grown SI 4H-SiC, $^{12}$C implanted SI 6H-SiC, and $^{12}$C implanted n-type 3C-SiC. **b.** ODMR spectrum of SI 4H-SiC as a function of $B$ parallel to the c-axis (upper) and at $B = 0$ (lower), showing six pairs of spin resonance lines. PL5 and PL6 appear faintly and are highlighted with dashed lines. **c.** ODMR spectrum of SI 6H-SiC as a function of c-axis $B$ (upper) and at $B = 0$ (lower), with a dashed lined highlighting the higher frequency QL5 resonance. **d.** ODMR spectrum of 3C-SiC as a function of [100]-oriented $B$ (upper) and at $B=0$ (lower). The 3C-SiC spins and the c-axis oriented defects in the hexagonal polytypes (PL1, PL2, PL6, QL1, QL2, and QL6) have $C_{3v}$ symmetry. The others (PL3, PL4, PL5, QL3, QL4, and QL5) are oriented along basal planes, resulting in the lower $C_{1h}$ symmetry and non-degenerate spin transitions at $B = 0$.

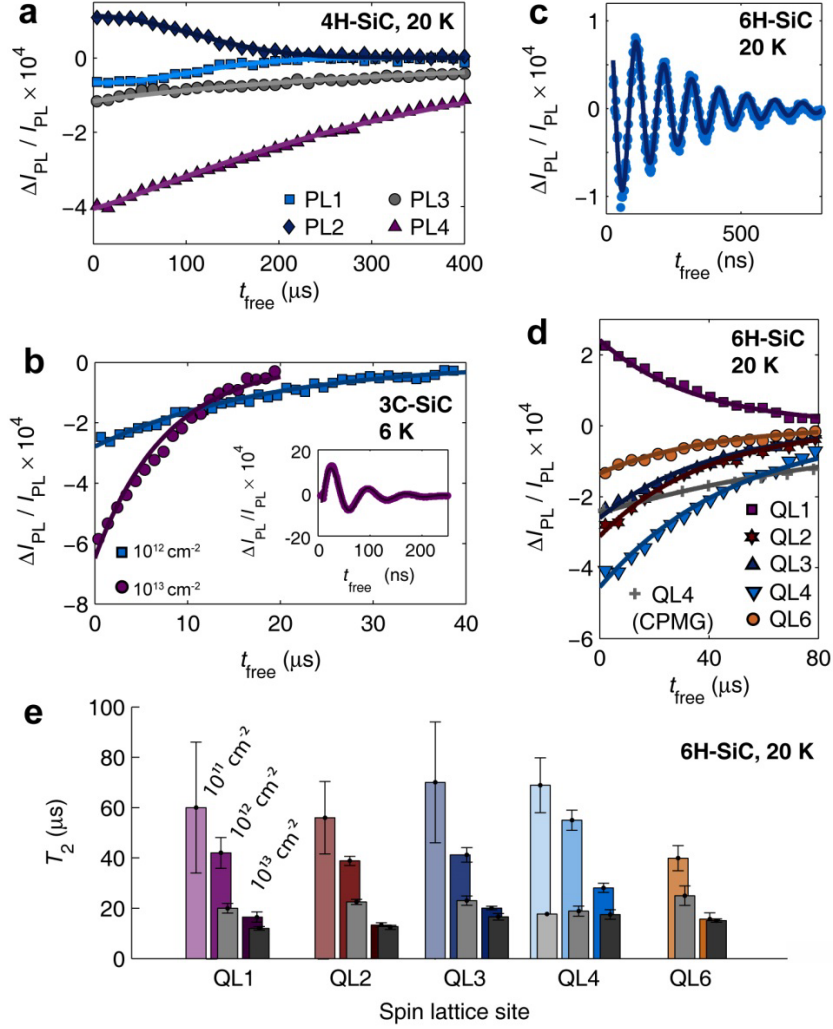

**Figure 2. Ensemble spin coherence at cryogenic temperatures. a.** Hahn-echo measurement of spin coherence of the neutral divacancies in as-grown SI 4H-SiC at 20K, showing $T_2$ times of $140 \pm 5$ μs (PL1), $144 \pm 3$ μs (PL2), $360 \pm 20$ μs (PL3), and $340 \pm 5$ μs (PL4). Long Rabi pulses (150 kHz) were used to minimize decoherence from instantaneous diffusion. **b.** Hahn-echo measurement of the 3C-SiC spins at 6K, implanted with $^{12}C$ at doses of $10^{12}$ cm$^{-2}$ and $10^{13}$ cm$^{-2}$, with respective decoherence times of $T_2 = 24 \pm 4$ μs and $T_2 = 12 \pm 1$ μs. Inset: Ramsey measurement of 3C-SiC spin dephasing, showing $T_2^* = 52 \pm 3$ ns. **c.** Ramsey measurement at 20 K of QL4 in SI 6H-SiC implanted with $^{12}C$ at a dose of $10^{12}$ cm$^{-2}$. RF pulses were detuned from resonance by 10 MHz. The fit is to an exponentially decaying sinusoid with $T_2^* = 250$ ns. **d.** Hahn-echo measurement for QL1-QL6 in $10^{12}$ cm$^{-2}$-implanted SI 6H-SiC at 20K, except for QL5, whose overlap with other ODMR lines at $B = 0$ inhibited zero-field measurements. CPMG dynamical decoupling for QL4 is also shown, with $T_{CPMG} = 106 \pm 2$ μs. The Rabi frequencies used (2.5 MHz) are less than the inhomogeneous linewidth, resulting in roughly half of the spins being driven. **e.** Comparison of 20 K Hahn-echo coherence times in n-type 6H-SiC (grey) and SI 6H-SiC (colored) for three different $^{12}C$ implantation doses and spin densities. The error bars are 95% confidence intervals from exponential fits to the Hahn echo data. $B=0$ for all the data in this figure.

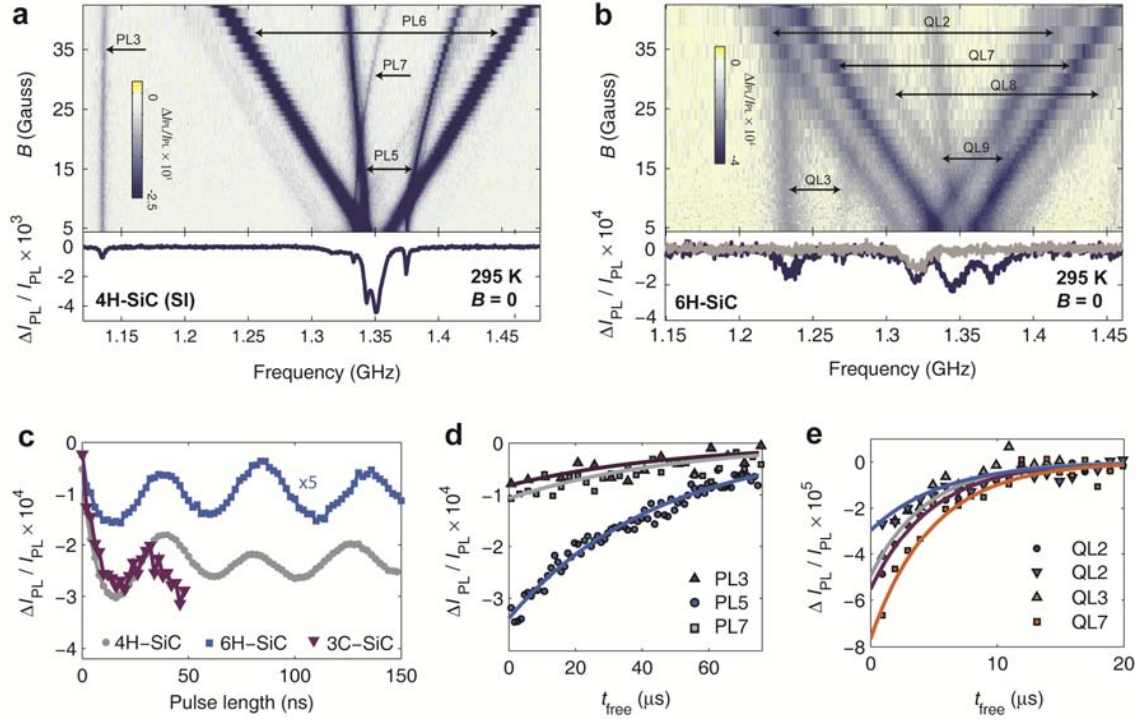

**Figure 3. Ensemble spin coherence at room temperature. a.** ODMR in as-grown SI 4H-SiC as a function of $B$ (upper) and at $B=0$ (lower). **b.** ODMR as a function of $B$ in n-type 6H-SiC (upper) and at $B=0$ (lower) for n-type (dark blue) and SI (grey) 6H-SiC, implanted at $10^{13}$ cm$^{-2}$ dose of $^{12}$C. QL7-QL9 have reduced $\Delta I_{PL}$ contrast as the temperature is lowered, disappearing by 200K. **c.** Rabi driving at room temperature: PL5 in SI 4H-SiC, QL2 in SI 6H-SiC (multiplied by a factor of ×5), and the 3C-SiC spin species. **d.** Hahn echo measurements of room temperature coherence for spin states in as-grown SI 4H-SiC. The fitted $T_2$ times are 50 ± 30 μs (PL3), 44 ± 2 (PL5), and 50±15 μs (PL7). **e.** Hahn echo measurement of room temperature coherence in n-type and SI 6H-SiC, implanted at the $10^{13}$ cm$^{-2}$ dose. The fitted $T_2$ times are 4.7 ± 0.6 μs (dark blue, QL2, n-type), 5.6 ± 1.8 μs (turquoise, QL2, SI), 4.4 ±1.5 μs (grey, QL3, n-type), and 4.9 ± 0.9 μs (purple, QL7, n-type). The uncertainties are 95% confidence intervals from exponential fits to the Hahn echo data.

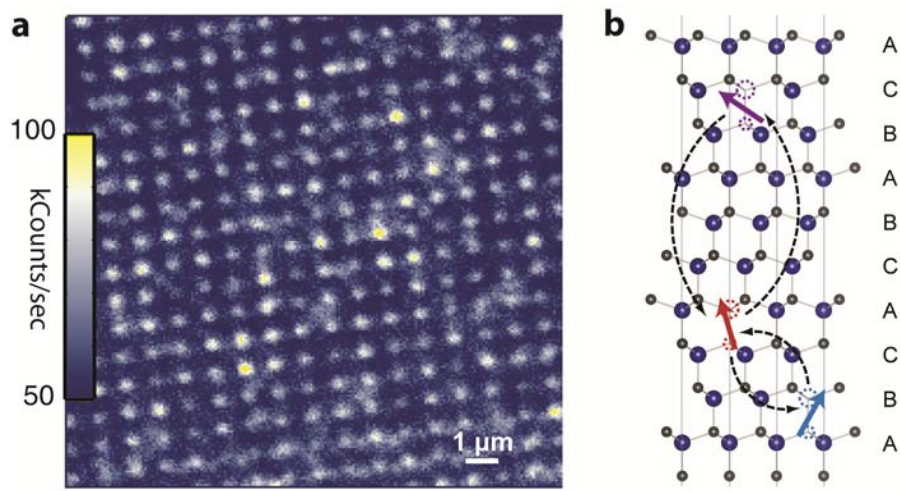

**Figure 4 | Patterned SiC spins and illustration of dipole-coupled spins. a.** Implanted spin ensembles in n-type 4H-SiC through a PMMA mask with 50 nm holes, using 10 keV-energy $^{12}$C ions at a $10^{13}$ cm$^{-2}$ dose. Additional characteristics of these implanted spins are given in Supplementary Figure S6. **b.** Illustration of dipole-coupled spin network in 6H-SiC in which each spin has a unique orientation and optical/ODMR signature. The three spins shown here are the (hh) divacancy (turquoise), ($k_1k_1$) divacancy (purple), and ($k_2k_2$) divacancy (red).

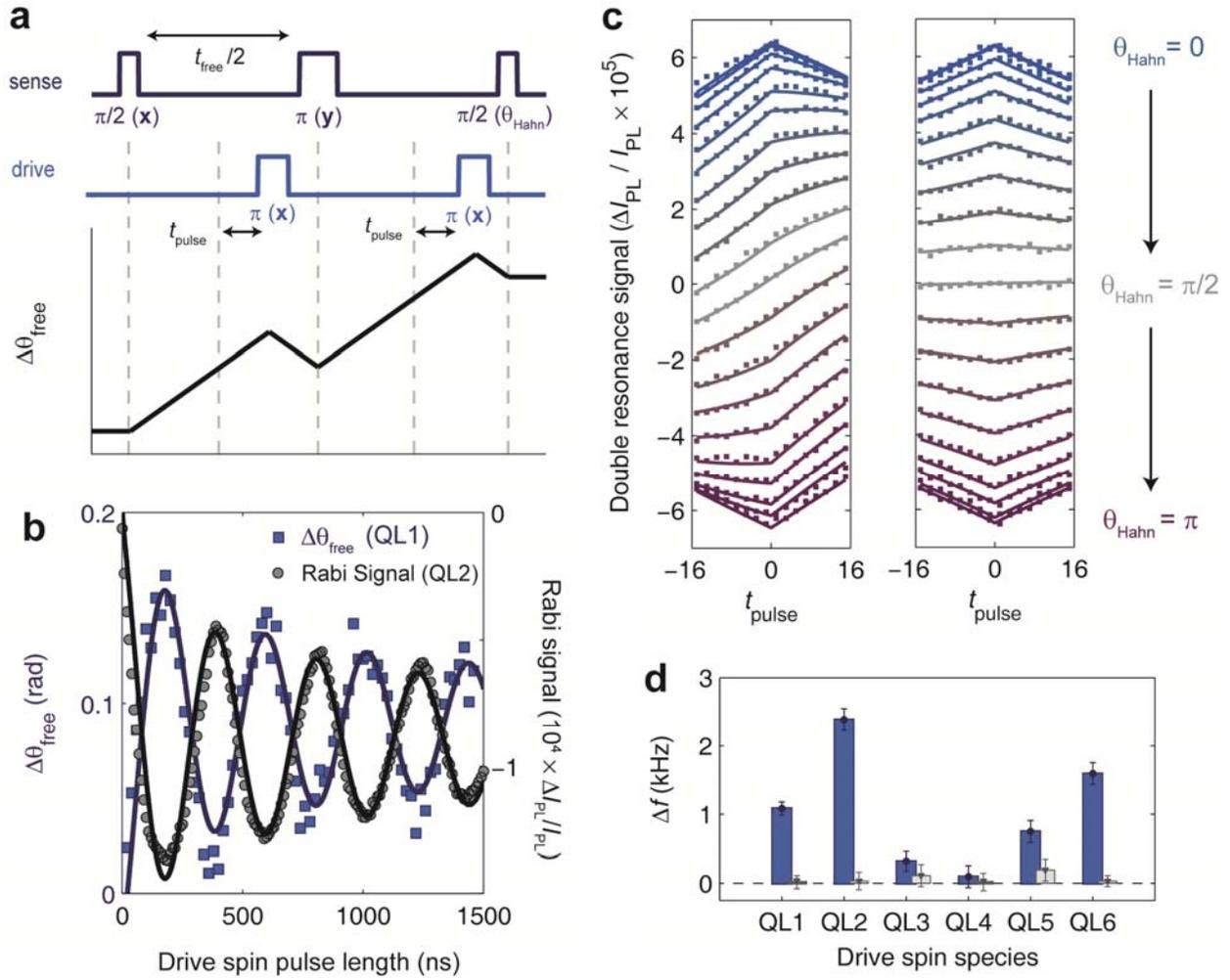

**Figure 5 | Magnetic dipole-coupled spin ensembles in 6H-SiC. a.** Pulse scheme for DEER measurements. As π pulses flip the orientation of the drive species spins, the sense species accumulates an extra phase ($\Delta\theta_{free}$) over a Hahn echo sequence. **b**. Varying the pulse duration of the drive spins (QL2) generates Rabi oscillations (right axis) and an oscillation in the DEER signal (left axis) corresponding to the sense spin species (QL1) acquiring $\Delta\theta_{free}$. The Rabi data are fitted according to an erfc-decaying sinusoid[30], and $\Delta\theta_{free}$ is fitted to a multiple of this function. **c.** DEER signal, applying two π pulses to the drive species and varying $\theta_{Hahn}$ and $t_{pulse}$. The left dataset is when the drive spins are polarized, and the right dataset is for unpolarized spins, for which $\Delta f$ vanishes. The data at different $\theta_{Hahn}$ are not artificially offset. The solid lines are global fits to Equation 1. **d.** Fitted $\Delta f$ for the six spin orientations, when the drive spin species is polarized (blue) and unpolarized (grey). The sense species is QL1 for all these data, except when QL1 is driven, in which case it is QL2. $B$ = 64 G and $t_{free}$ = 32 μs, and $T$ = 20 K for all the data in Fig. 5.

# Supplementary Information for "Polytype Control of Spin Qubits"

Abram L. Falk[1], Bob B. Buckley[1], Greg Calusine[1], William F. Koehl[1], Viatcheslav V. Dobrovitski[2], Alberto Politi[1], Christian A. Zorman[3], Philip X.-L. Feng[3], David D. Awschalom[1]

*1. Center for Spintronics and Quantum Computation, University of California, Santa Barbara, California 93106, USA*

*2. Ames Laboratory and Iowa State University, Ames, Iowa 50011, USA*

*3. Department of Electrical Engineering, Case School of Engineering, Case Western Reserve University, Cleveland, OH 44016, USA*

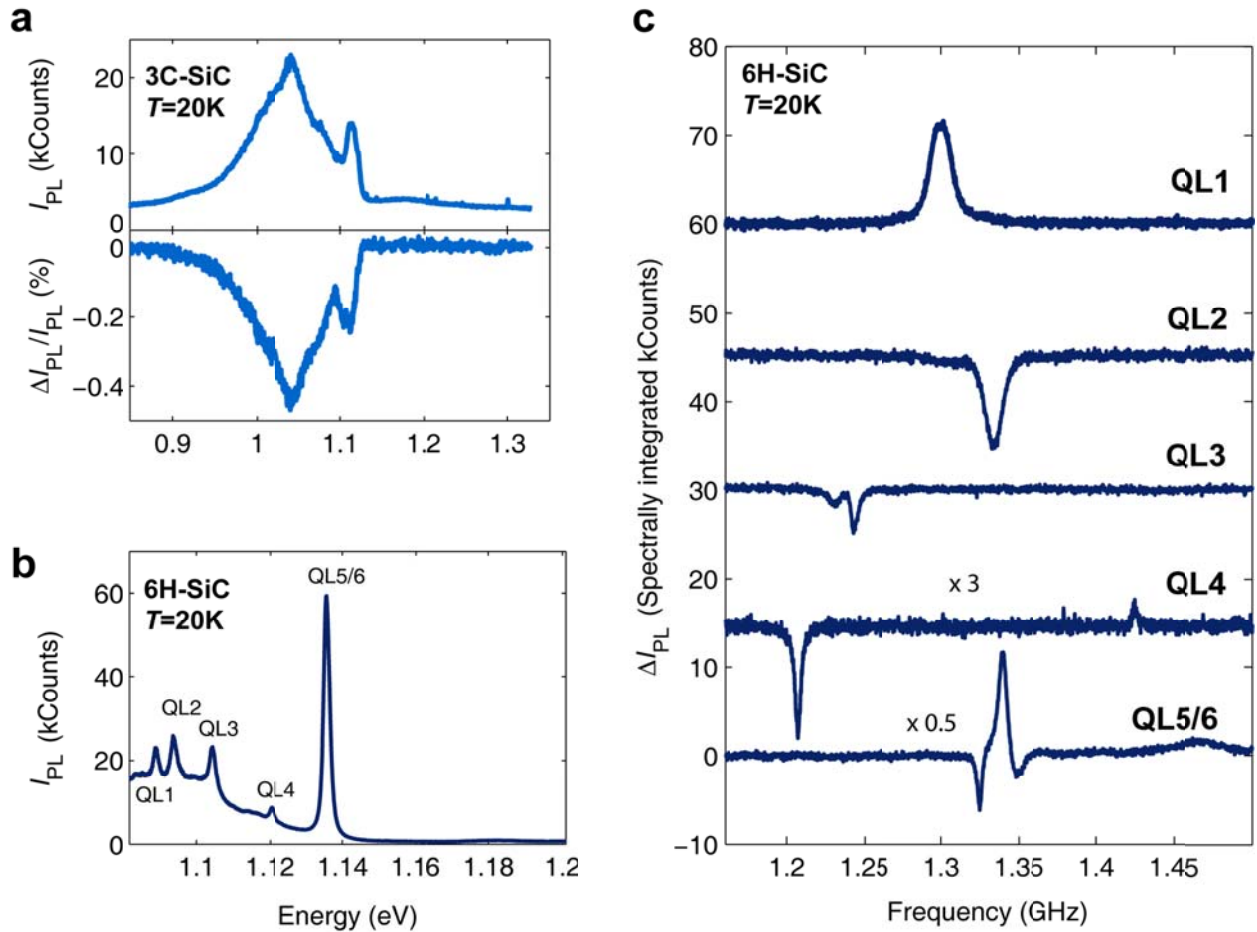

**Supplementary Figure S1. Association of photoluminescence and optically detected magnetic resonance peaks in 6H- and 3C-SiC**. (**a**) Top: a photoluminescence (PL) spectrum typical of a $^{12}$C implanted, unintentionally n-type 3C SiC epilayer. The PL spectrum consists of the zero-phonon line (ZPL) at 1.12 eV and a phonon sideband that peaks around 1.05 eV. Bottom: the spectrally resolved change in PL ($\Delta I_{PL}$) when a 1.328 GHz microwave signal is applied to the sample. The shape of $\Delta I_{PL}$ matches $I_{PL}$, except for the sign. A frequency independent feature originating from the silicon substrate was subtracted from the optically detected magnetic resonance (ODMR) spectrum so as not to obscure the ZPL. (**b**) PL spectrum of 6H-SiC. (**c**) ODMR spectra in 6H-SiC corresponding to the 6H-SiC ZPLs. By spectrally narrowing $I_{PL}$ to each zero-phonon line, we could then measure the wavelength dependent $\Delta I_{PL}$ when RF power is applied and associate each ODMR resonance to a PL peak. The magnetic field ($B$) dependence (not shown) of the ODMR spectrum corresponding to the PL peak at 1.135 eV clearly shows both a c-axis oriented defect and a basal-oriented defect. This observation indicates that the PL peak at 1.135 eV is two peaks lying on top of each other, which we label QL5 and QL6. This identification of two overlapping peaks is consistent with previous PL measurements[23] of the UD-2 spectrum in 6H-SiC, which have found two closely-spaced zero-phonon lines at 1.135 eV.

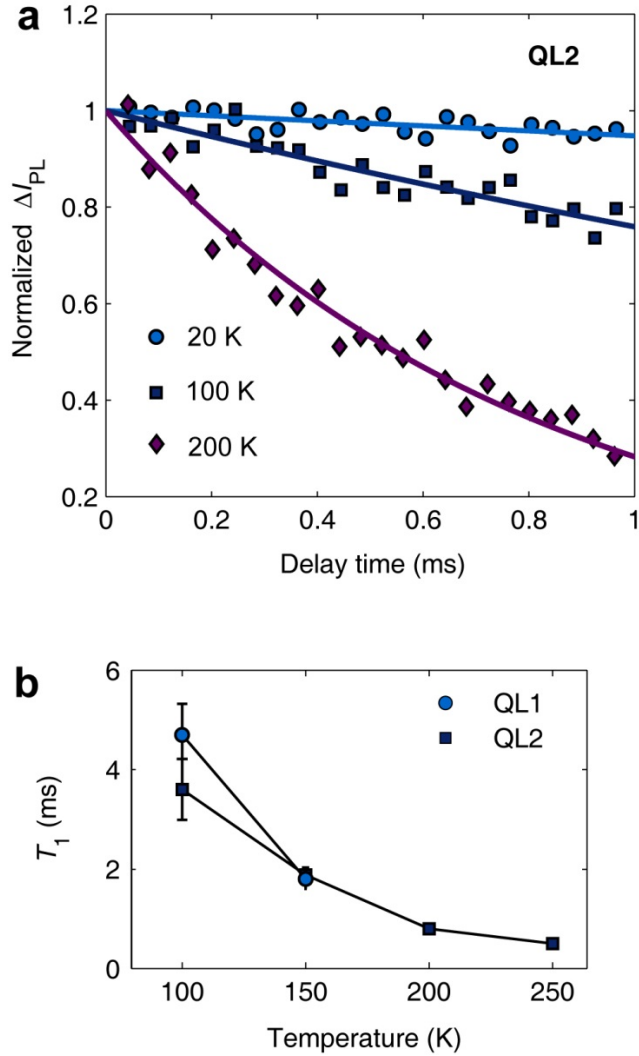

**Supplementary Figure S2. Measurement of spin relaxation times in 6H-SiC. (a)** Measurement of spin relaxation times ($T_1$ times) for QL2 at temperatures ranging from 20 K to 250 K. **(b)** Temperature dependence of $T_1$ for QL1 and QL2. To perform these spin relaxation measurement sequences, we optically polarized the spins, rotated a spin species with an RF $\pi$ pulse immediately after the optical pulse, waited a variable delay time, and then optically measured the spin projection after that delay. The $\Delta I_{PL}$ signal is locked into the presence or absence of the RF pulse at 20 Hz. As the spins relax at long delay times, $\Delta I_{PL}$ decays. For QL2 in SI 6H-SiC, $T_1$ is 620 ± 100 µs at 250 K and increases to 5 ± 1 ms as the sample is cooled to 100K. At 20K, the $T_1$ times for QL1, QL2, QL3, and QL4 were respectively 24 ± 4 ms, 18 ± 6 ms, 8 ± 1 ms, and 11 ± 3 ms.

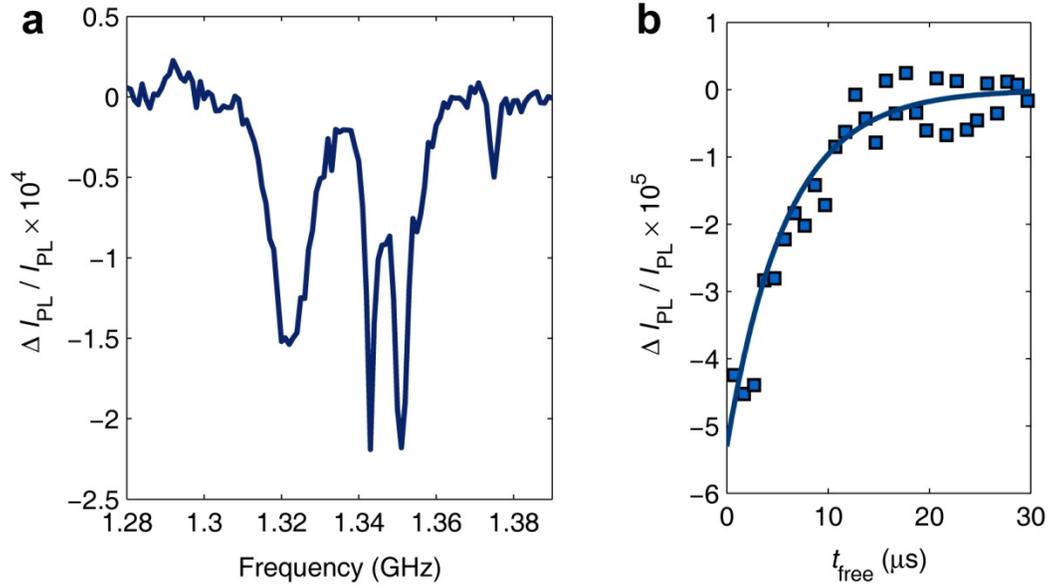

**Supplementary Figure S3. Room-temperature spin coherence of PL1 in implanted 4H-SiC. (a)** Room-temperature ODMR spectrum for semi-insulating 4H-SiC that is implanted with a $10^{13}$ cm$^{-2}$ dose of $^{12}$C, showing an additional ODMR line at 1.322 GHz that is not in the as-grown material (Figure 3a in the main text). This ODMR line corresponds to PL1. **(b)** Hahn-echo coherence measurement of PL1, showing a $6 \pm 1$ μs spin coherence time ($T_2$ time). The shorter $T_2$ time in this implanted sample relative to the $T_2$ times in the as-grown semi-insulating 4H-SiC material (~50 μs at room temperature) is likely due to material damage from the ion implantation. Implanted n-type 4H-SiC also exhibited a PL1 ODMR line at room temperature.

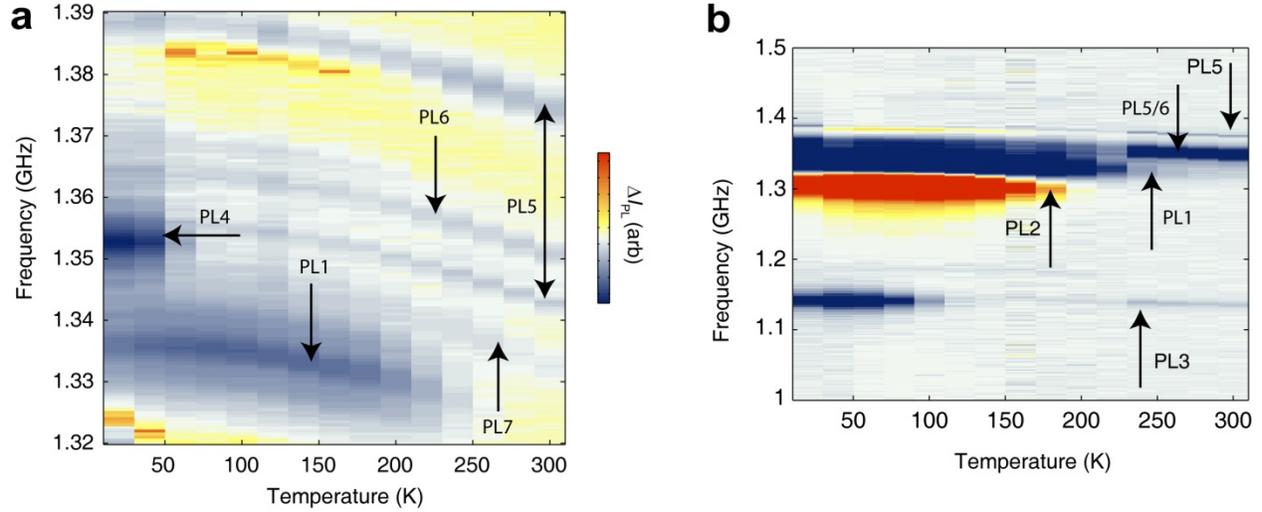

**Supplementary Figure S4. Temperature dependence of ODMR in 4H-SiC.** (**a**) ODMR spectrum of as-grown 4H-SiC in a narrow frequency range, at $B$=0. (**b**) Wider frequency range to include PL3 at 1.14 GHz, also at $B$=0. The wide ODMR lines are due to a saturated color scale. In 4H-SiC, all of the ODMR lines that we observe uniformly shift 14 MHz lower in energy when warmed from 20K to room temperature. The spin-dependent optical contrast $\Delta I_{PL}/I_{PL}$ is affected as the sample temperature ($T$) is increased, and many of the ODMR lines can no longer be observed when the sample is at room temperature. However, an ODMR line corresponding to PL3 remains visible at room temperature, as does PL1 when the substrate is implanted with $^{12}$C ions and annealed. The ODMR features corresponding to PL5 and PL6 are also visible at room temperature, and their $\Delta I_{PL}/I_{PL}$ values are not strongly affected by warming the sample from cryogenic to room temperature. Unlike the neutral divacancies in 4H-SiC (PL1-PL4), PL5 and PL6 are not conclusively identified.

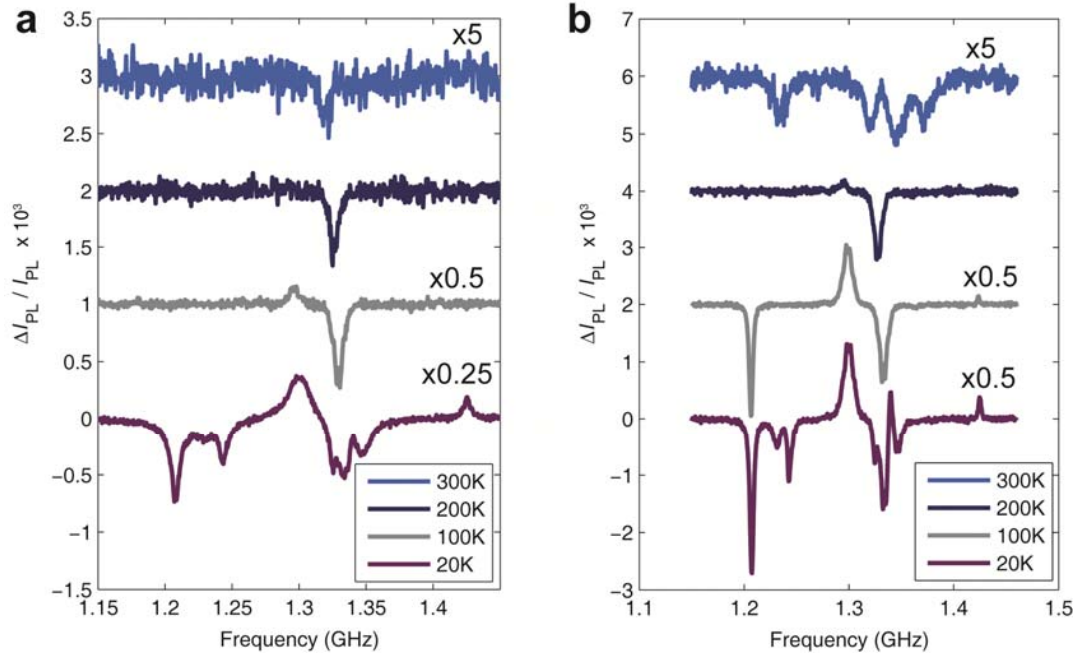

**Supplementary Figure S5. Temperature dependence of ODMR in 6H-SiC.** The $^{12}$C implanted semi-insulating and n-type substrates show the same ODMR lines at 20K (QL1-QL6). However, at room temperature, the implanted n-type 6H-SiC substrate and the implanted semi-insulating 6H-SiC differ, with the n-type substrate exhibiting several ODMR lines (QL7-QL9) that are not in the semi-insulating substrate. **(a)** The only ODMR feature in semi-insulating 6H-SiC at room temperature is the QL2 line at 1.32 GHz. **(b)** In implanted n-type 6H-SiC, there are several new ODMR features that arise at higher temperatures. At lower temperatures, QL7-QL9 lose their $\Delta I_{PL}/I_{PL}$ contrast.

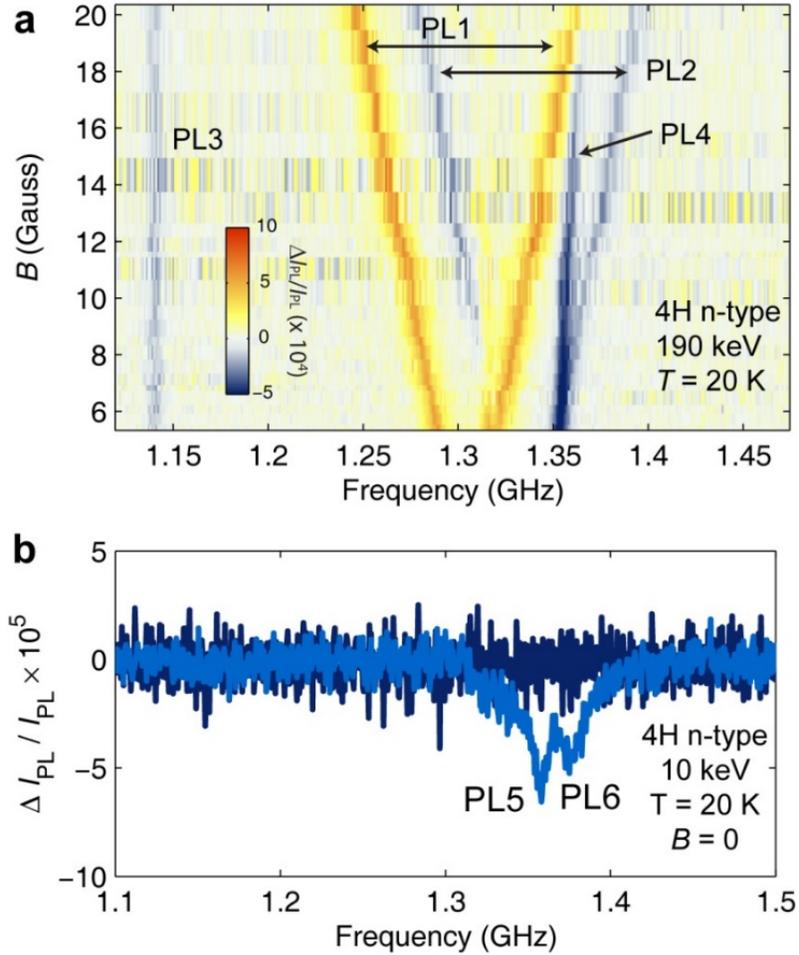

**Supplementary Figure S6. ODMR spectrum of $^{12}$C-implanted n-type 4H-SiC.** (**a**) *B*-field dependent ODMR spectrum when the implantation energy is 190 keV and the dose is $10^{13}$ cm$^{-2}$. ODMR lines corresponding to PL1-PL4 are exhibited. Even when implanted with $^{12}$C ions and annealed, the n-type 4H-SiC had the smallest ODMR signal of any of the hexagonal polytype samples that we measured. However, it also exhibited the smallest amount of unwanted background luminescence under photoexcitation. For this reason, we chose it for our demonstration of patterned ion implantation. (**b**) ODMR spectrum at *B*=0 of the patterned ion-implanted n-type 4H-SiC, for the sample shown in Figure 4a in the main text. Implantation at a lower energy (10 keV) was used for this sample so that the ions would not penetrate the PMMA mask used for the patterning[31]. ODMR spectra measured at 20K in both the masked (dark blue) and unmasked (turquoise) regions of that sample are shown. One of the peaks is c-axis oriented and the other is basal oriented, corresponding to PL5 and PL6 respectively.

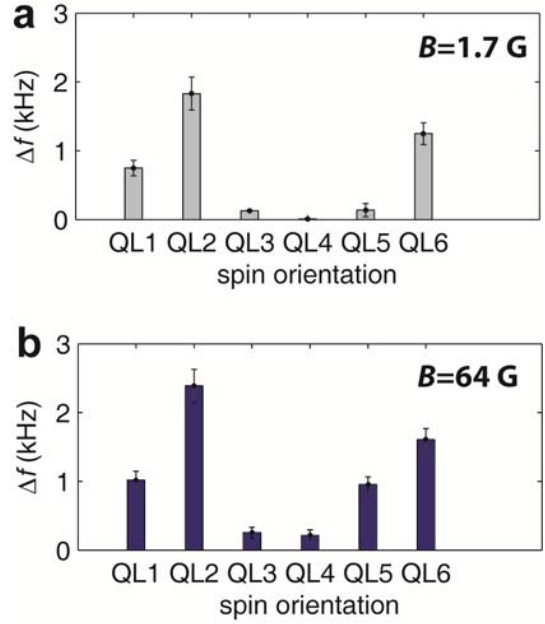

**Supplementary Figure S7. Magnetic field dependence of ensemble spin coupling strength** at **(a)** $B$=1.7 G and **(b)** $B$=64 G. For these data, $t_{\text{free}}$ = 32 μs and $T$ = 20 K. The data in **(b)** are the same as in Figure 5 in the main text but reprinted here to be compared to those in **(a).** While $\Delta f$ for the c-axis oriented defects (QL1, QL2, and QL6) is roughly the same at the low (1.7 G) and high (64 G) fields, the basal-oriented defects QL4 and QL5 exhibit much stronger $\Delta f$ values (greater than a factor of 5) at the higher field. This can be understood by the fact that, when they are modeled by equation S1, their Hamiltonians have large $E$ terms and their eigenstates have low magnetic moments at $B$=0. As $B$ increases, the magnetic moments of the eigenstates increase. QL3 is also basal oriented, but the relatively small E term in its Hamiltonian causes $\Delta f$ to be less dependent on $B$.

| PL line | Other name | Identification | Orientation | Optical energy (eV) | ODMR 1 (GHz) (*) | ODMR 2 (GHz) (*) | D (GHz) | E (MHz) | Room temp? | 20K? |
|---|---|---|---|---|---|---|---|---|---|---|
| | | | | 4H-SiC | | | | | | |
| PL1 | P6b | (hh) divacancy | c-axis | 1.095 | 1.336 | - | 1.336 | < 1 | implanted | yes |
| PL2 | P6'b | (kk) divacancy | c-axis | 1.096 | 1.305 | - | 1.305 | < 1 | no | yes |
| PL3 | P7'b | basal divacancy | basal | 1.119 | 1.140 | 1.304 | 1.222 | 82.0 | yes | yes |
| PL4 | P7b | basal divacancy | basal | 1.150 | 1.316 | 1.353 | 1.334 | 18.7 | no | yes |
| PL5 | | unknown | c-axis | 1.189 | 1.356 | 1.389 | 1.373 | 16.5 | yes | yes |
| PL6 | | unknown | basal | 1.194 | 1.365 | - | 1.365 | < 1 | yes | yes |
| PL7 | | unknown | basal | - | 1.333 | - | - | - | yes | yes |
| | | | | 6H-SiC | | | | | | |
| QL1 | | unknown | c-axis | 1.088 | 1.300 | - | 1.300 | < 2 | no | yes |
| QL2 | | unknown | c-axis | 1.092 | 1.334 | - | 1.334 | < 2 | yes | yes |
| QL3 | | unknown | basal | 1.103 | 1.228 | 1.243 | 1.236 | 7.5 | yes | yes |
| QL4 | | unknown | basal | 1.119 | 1.208 | 1.425 | - | 109 | no | yes |
| QL5 | | unknown | basal | 1.134 | 1.340 | - | 1.383 | - | yes | yes |
| QL6 | | unknown | c-axis | 1.134 | 1.347 | - | 1.347 | < 5 | no | yes |
| QL7 | | unknown | c-axis | - | 1.345 (**) | - | 1.345 (**) | < 5 | yes | no |
| QL8 | | unknown | c-axis | - | 1.371 (**) | - | 1.371 (**) | < 5 | yes | no |
| QL9 | | unknown | basal | - | 1.332 (**) | 1.365 (**) | 1.349 (**) | 17 (**) | yes | no |
| | | | | 3C-SiC | | | | | | |
| RL1 | Ky5 | unknown | [111] | 1.120 | 1.328 | | 1.328 | < 2 | substrate-dependent | yes |

* At B=0 and at 20K, unless not visible at low temperature, in which case the room-temperature values are given

** Room temperature frequency

**Supplementary Table S1. Observed defect optical energies, ODMR lines, and orientations.** The neutral divacancies in 4H-SiC are spin 1, $g=2$ spins.[9] The c-axis defects have $C_{3v}$ symmetry, and the basal defects have the lower $C_{1h}$ symmetry. They can be modeled with the Hamiltonian:

$$H = \mu_B g \mathbf{B} \cdot \mathbf{S} + D\left(S_Z^2 + \frac{1}{3}S(S+1)\right) + E(S_X^2 - S_Y^2) \qquad (S1)$$

where $\mu_B$ is is the Bohr magneton, $g$ is the Lande g-factor, **B** is the external magnetic field, **S** is the spin, and $D$ and $E$ are the axially symmetric and anisotropic components of the crystal field interaction, respectively. At $B=0$, the spin transition energies are $D-E$ and $D+E$. In 6H-SiC, the photoluminescence (PL) and optically-detected magnetic resonance (ODMR) transition energies of QL1-QL6 are similar to the neutral divacancies (PL1-PL4) in 4H-SiC, suggesting a close association. However, they are not conclusively identified.

The PL lines corresponding to QL1-QL6, which have been seen before,[23] are known as the UD-2 spectrum in 6H-SiC and are attributed to complexes involving intrinsic defects. However, no previous work has associated defects with the $D$ and $E$ values that we observe with the QL1-QL6 PL lines. Son et al.[20] report observations of ODMR with some of the zero-field splittings that we observe, but they do not observe a spectral dependence for the ODMR signal. Lingner et al.[19] observe some similar zero-field splittings to those of our 6H-SiC defects. However, the *D* and *E* values of their defects are not exactly the same as ours, and they associate their defects with lower energy optical transitions.

In 3C-SiC, a previous PL band similar to the one we observe has been observed[22] but associated with a spin-1/2 ODMR signal. Another study[21] has associated this band with a spin-1 ODMR signal but observed a zero-field splitting 50-80 MHz different than the one we observe and did not determine a spectral dependence for this ODMR signal. The PL bands from both of these reports and the spin-1 ODMR signal observed depend on the sample annealing temperature in a manner similar to our observations. An electron paramagnetic resonance study[14] has observed a spin-1 defect in 3C-SiC with zero-field splitting and symmetry similar to the one we observe, which persists up to room temperature. It tentatively attributed this signal to a neutral divacancy.

**Supplementary Note 1. Estimation of defect creation efficiency**

For the 190 keV energy $^{12}$C implantations used in the main text, we estimate that we have created defects in 6H-SiC with a 5% creation efficiency, which we define here as the number of fluorescent defects of a single species created per implanted 190 keV $^{12}$C ion. To arrive at that estimation, we put together three factors:

1. For the three c-axis oriented defect species in 6H-SiC, our double resonance measurements show that the density of implanted spins was approximately $10^{16}$ cm$^{-3}$.

2. Those chips were implanted with $^{12}$C ions at a dose of $10^{13}$ cm$^{-2}$.

3. Finally, by modeling the ion implantation process using transport range of ions in matter (TRIM) simulations, we find that the vacancies generated by the 190 keV ion implantation process are concentrated in the upper 400 nm of the chip. For these simulations, used the TRIM software package created by James Zieglier, documented at http://www.srim.org.

Putting together these three factors, we find a 5% creation efficiency for 190 keV $^{12}$C ions. Because a single implanted $^{12}$C ion causes a cascade of damage in the crystal, the creation efficiency of fluorescent defects per crystal vacancy is significantly lower. For 190 keV implanted ions, our TRIM calculations model 620 vacancies per implanted ion. For the 10 keV energy ions $^{12}$C ions, TRIM calculations model 130 vacancies per implanted ion.

**Supplementary Note 2. Experimental considerations and calculation of optical polarization for the double electron-electron resonance measurements**

The double electron-electron resonance (DEER) measurements presented here and in the main text all use a hardware lock-in technique based on IQ modulation. Defining the Bloch sphere such that the sense species spins are initialized in the **+z** direction, the phase of the first π/2 pulse of the Hahn echo sequence rotates the spin around the **x** axis. The π pulse rotates the spin around the **y** axis. The phase of final π/2 pulse is modulated between being a rotation around the $\theta_{Hahn}$ axis and the $-\theta_{Hahn}$ axis, where $\theta_{Hahn}$ is the angle of a vector in the x-y plane, defined relative to the x-axis.

The optical initialization/readout time is 100 μs, using a 975 nm laser focused to roughly 20 μs by a 14 mm lens with 60 mW of power reaching the sample. This time was chosen to match or exceed the saturation time of $\Delta I_{PL}$ with laser pulse length, so that the spins would be polarized at the start of each cycle. The timescale of this saturation curve is highly dependent on the optics, such as the lens choice at the sample and whether there is a pinhole at the detector. For the DEER measurements presented here, we used a 35 mm lens to focus the collimated fluorescence onto a 300 μm detector with no pinhole, and the measured $\Delta I_{PL}$ saturation curve showed a 30 μs saturation time. Since there is no pinhole, these optics are not confocal. For confocal optics, the timescale of optical readout is much faster, but the signal strength is much weaker due to the smaller volume of PL collection.

Because of the long readout/initialization time, the spatial profile of the optically initialized (magnetized) spins is large-radius disk with a decaying tail outside the disk, whose radius is significantly larger than the confocal laser spot size. The diameter of this disk is not precisely known but it is estimated to be in the 100 μm range. Based on stopping range of ions in matter simulations (SRIM), we estimate that the thickness of this disk is 400 nm.

In order to calculate the defect optical polarization, we first use the results of our DEER measurements to separately calculate the defect density and the bulk magnetization from those defects. By applying equation 1 in the main text to the DEER data, we can use the fitted τ to infer the spin density by applying a statistical model of instantaneous diffusion[26,27,32] (see Supplementary Note 3).

For a homogeneous distribution of spins and small exchange coupling, the free induction decay of the DEER magnitude is given by:

$$I_{DEER}(t) = e^{-kCFt}, \quad (S2)$$

where

$$k = \frac{2\pi\mu_0 g^2 \mu_B^2}{9\sqrt{3}\hbar}, \quad (S3)$$

$C$ is the spin concentration, and $F$ is the net fraction of spins flipped by the drive pulses.

The factor $F$ is used to account for the imperfect Rabi pulses used in the measurement sequence. To calculate $F$, we consider that the carrier-frequency-dependent probability of a spin flip given by our π pulses is given by Rabi's formula:

$$p_{\text{flip}}(\Omega, \delta, t) = \frac{\Omega^2}{\Omega^2 + \delta^2} \sin^2\left(\frac{1}{2}\sqrt{\Omega^2 + \delta^2} 2\pi t\right) \quad (S4)$$

where $\Omega$ is the Rabi frequency, δ is the detuning from the center frequency of the pulse and $t$ is the pulse length. The total fraction of spin polarization excited by a π pulse is given by integrating $p_{\text{flip}}$ weighted by the low-power ODMR spectrum $\rho(\omega)$, and finding the maximum of this integral as a function of pulse length. This fraction is given by:

$$F = \max_t \left( \frac{\int p_{\text{flip}}(\Omega, \omega - \omega_0, t)\rho(\omega)d\omega}{\int \rho(\omega)d\omega} \right), \quad (S5)$$

Where $\omega_0$ is the carrier frequency. Note that this maximum occurs near but not exactly at $t=1/(2\Omega)$, due to unequal rotation rates of defects as a function of detuning.

For the c-axis spins and $\Omega = 2.5$ MHz, as used in Figure 5 of the main text, we calculate $F$ to be 0.51, 0.56, and 0.57 for QL1, QL2, and QL6 respectively. Our DEER measurements use two $\pi$ pulses, and the net change in polarization of the drive species between the optically initialized state and its state after the first $\pi$ pulse can be different from the net change between the first and second $\pi$ pulses. However, such a difference would manifest itself as a $t_{\text{pulse}} \to -t_{\text{pulse}}$ asymmetry in the DEER data (Figure 5c) at $\theta_{\text{Hahn}} = 0$ and $\theta_{\text{Hahn}} = \pi$. While such asymmetry is not zero, it is clearly small, and we neglect it for this analysis.

To calculate the spin density, we then use $\tau$ values extracted from fits to equation 1 in the main text, which are $\tau = 130 \pm 15$ μs, $\tau = 96 \pm 3$ μs, and $\tau = 150 \pm 30$ μs for QL1, QL2, and QL6 respectively, when the drive spin baths are unpolarized. Using $C = 1/(\tau k F)$, we find that the concentration of c-axis spins is $9.1 \times 10^{15}$ cm$^{-3}$, $1.1 \times 10^{16}$ cm$^{-3}$, and $7.1 \times 10^{15}$ cm$^{-3}$ for QL1, QL2, and QL6 respectively. We did not calculate the density of basal-oriented spins, because of high uncertainties in the DEER measurements for these species.

To calculate the optical polarization, we use $\Delta f$ fitted in equation 1 of the main text to infer the average change in magnetic field magnitude when the drive spins are flipped by a $\pi$ pulse. We find these values to be $\Delta f = 1.1 \pm 0.1$ kHz, $2.4 \pm 0.1$ kHz, and $1.6 \pm 0.2$ kHz for QL1, QL2 and QL6 respectively. Using the concentrations calculated above, we can then compare the value of flipped magnetization to the spin density. The change in magnetic field corresponding to a precession frequency change of $\Delta f$ can be calculated from the Zeeman splitting:

$$\Delta B = \frac{h\Delta f}{g\mu_B}. \quad (S6)$$

Assuming that the magnetization profile is a large-radius magnetized disk, the magnetization change due to a $\pi$ pulse is then simply $\Delta B/\mu_0$. Finally, the optical polarization ($P_{\text{optical}}$) can be calculated as the percentage of spins that must be originally magnetized to achieve this magnetization. For a $\Delta m_s = 1$ change in spin:

$$P_{\text{optical}} = \frac{\Delta B}{C \cdot F \mu_0 (g\mu_B)} = \frac{4\pi^2}{9\sqrt{3}} \Delta f \cdot \tau \quad (S7)$$

Using equation S7 and the experimentally fitted $\Delta f$ and $\tau$ values, we find the degree of optical polarization of the three c-axis spins. This leads us to calculate the degree of optical polarizations to be 36%, 58%, and 60% for QL1, QL2, and QL6 respectively. As noted in the main text, since PL is collected from spins that are not in the uniformly polarized center of the optical excitation area, these values are likely to be lower bounds for the perfectly polarized values.

**Supplementary Note 3. Theoretical considerations for double electron-electron resonance analysis.**

**A. Effect of strong spin polarization – qualitative discussion.**

**B. Local field distribution in the presence of strong spin polarization.**

These sections are not available in the arXiV version of this article. They are available in the peer-reviewed *Nature Communications* article: A. L. Falk et al, *Nat. Commun.*, **4**, 1819 (2013): doi:10.1038/ncomms2854

**Supplementary References**